\documentclass[
 aip,
 amsmath,amssymb,
 reprint
]{revtex4-1}

\usepackage{graphicx}% Include figure files
\usepackage{dcolumn}% Align table columns on decimal point
\usepackage{bm}% bold math
\usepackage{tabularx}

\usepackage[utf8]{inputenc}
\usepackage[T1]{fontenc}
\usepackage{etoolbox}
\usepackage{lipsum}
\usepackage{xcolor}
\usepackage{soul}
\usepackage[normalem]{ulem}

\newtheorem{theorem}{Theorem}
\newtheorem{lemma}[theorem]{Lemma}

\newcommand{\x}{\mathbf{x}}
\newcommand{\im}{\mathrm{im}}
\newcommand{\XX}{\mathbf{X}}
\newcommand{\A}{\mathcal{A}}
\newcommand{\RR}{\mathbb{R}}
\newcommand{\K}{\mathcal{K}}
\renewcommand{\P}{\mathcal{P}}
\renewcommand{\L}{\mathcal{L}}
\newcommand{\B}{\mathcal{B}}
\newcommand{\N}{\mathcal{N}}
\newcommand{\I}{\mathcal{I}}
\newcommand{\DD}{\mathbb{D}}

\newcommand{\rh}{\hat{\rho}}
\newcommand{\at}{\tilde{a}}

\newcommand{\sign}{\mathrm{sign}}

\makeatletter
\def\@email#1#2{%
 \endgroup
 \patchcmd{\titleblock@produce}
  {\frontmatter@RRAPformat}
  {\frontmatter@RRAPformat{\produce@RRAP{*#1\href{mailto:#2}{#2}}}\frontmatter@RRAPformat}
  {}{}
}%
\makeatother
\begin{document}

%\preprint{AIP/123-QED}

\title{Controlled transport of fluid particles by microrotors in a Stokes flow using linear transfer operators} 
% Force line breaks with \\
\author{Jake Buzhardt}

\author{Phanindra Tallapragada}
\email{jbuzhar@g.clemson.edu, ptallap@clemson.edu}
 \affiliation{Department of Mechanical Engineering, Clemson University, Clemson 29634, SC, USA}%

\date{\today}

\begin{abstract}
The manipulation of a collection of fluid particles in a low Reynolds number environment has several important applications. As we demonstrate in this paper, this manipulation problem is related to the scientific question of how fluid flow structures direct Lagrangian transport. We investigate this problem of directing the transport by manipulating the flow, specifically in the Stokes flow context, by controlling the strengths of two rotors fixed in space. We demonstrate a novel dynamical systems approach for this problem and apply this method to several scenarios of Stokes flow in unbounded and bounded domains.  
Further, we show that the time-varying flow field produced by the optimal control can be understood in terms
of dynamical structures such as coherent sets that define Lagrangian transport.
We model the time evolution of the fluid particle density using finite dimensional approximations of the Liouville operators for the micro-rotor flow fields.  Using these operators, the particle transport problem is framed as an optimal control problem, which we solve numerically.  This framework is then applied to the problem of transporting a blob of fluid particles in domains with different boundary conditions: free space, near to a plane wall, in a circular confinement, and the transport of two distributions of particles to a common target.  These examples demonstrate the effectiveness of the proposed framework and also shed light on the effects of boundaries on the ability to achieve a desired fluid transport using a rotor-driven flow.
\end{abstract}

\maketitle

\section{Introduction}\label{sec:intro}
Understanding and controlling the motion of fluid particles in the low Reynolds number regime has become an increasingly significant problem in recent years, particularly in the realm of microrobotics and microfluidics.  Microrotors and micropumps propelled by various mechanisms have been proposed as a useful means of transporting fluid particles or other submersed cargo in a microfluidic solution \cite{iverson2008recent,zhou2016chemistry,xu2017fuel,wang2018micropumps}.  In this paper, we develop a method based on recent advances in data-driven dynamical systems to model and control the transport of distributions of fluid particles using microrotors in a Stokes flow. 

This work is largely motivated by recent advances in micro-robotics, where artificial microscale swimming robots have been developed which are able to achieve locomotion in the micro-fluidic setting.  The proposed applications of such microrobots are typically biomedical applications\cite{sitti_IEEE_2015,ceylan2017microrobots_bioengineering,bente2018biohybrid}, including targeted therapeutics and drug delivery\cite{park2017multifunctional, ceylan20193d}, assisted fertilization\cite{medina2016cellular}, minimally invasive microsurgery\cite{edd2003biomimetic, roy2006mems}, and cell sorting and manipulation\cite{ren20193d,chen20232d}, among many others.  For many of these applications, the controlled transport of fluid particles, drugs, capsules, cells, or other cargo is necessary.  Using a micro-robot to manipulate such cargo to a target destination has been proposed in many previous works\cite{ren20193d,chen20232d,steager2013automated,tottori2012magnetic}, including by the authors\cite{buzhardt2021magnetically}, where it was suggested that such swimmers can be used stir the fluid to transport cargo in a contactless manner.  In this work, we take a useful step towards solving these complex manipulation problems by showing that distributions of fluid particles can be effectively controlled using the fluid velocity fields produced by a pair of micro-rotors by modulating the torques of these rotors.

In recent decades, significant attention has been given to the application of dynamical systems theory to problems of fluid transport at low Reynolds number, with much of this attention focusing on mixing by chaotic advection \cite{aref1984stirring,aref2017frontiers,ottino1989mixing}. 
Such research has been largely inspired by applications including industrial mixing, design of microfluidic lab-on-a-chip devices, and biomedical applications such as cell sorting and targeted drug delivery.  
While mixing is important to many such applications, many also require an ability to transport packets of fluid or a concentrated, passive scalar in a controlled way to a target destination while minimally mixing, stretching, or distributing the blob.  
Despite its growing practical importance, this area has received considerably less research attention. 
In particular, the application considered in this work is motivated by recent works which have shown that biological and artificial microswimmers can generate significant transport of passive fluid particles\cite{jin2021collective,dunkel2010swimmer,mueller2017fluid,shum2017entrainment} and finite-sized particles in the fluid\cite{vaccari2018cargo,purushothaman2021hydrodynamic,mathijssen2018universal}, and the potential usefulness of this observation for the design of micro-scale robots for controlled fluid or cargo transport \cite{uchida2010synchronization,huang2014generating,ye2014dynamic,matsunaga2019controlling,buzhardt2021magnetically}.  

In this work, we study the problem of steering an ensemble of fluid particles in a Stokes flow from an initial particle distribution to a final distribution, where the particles are advected by the flow field generated by a pair of fixed rotors.  
In our formulation, the distribution of fluid particles is described by a density function, and a data-driven method based on a finite dimensional approximation of the Liouville operator associated with the rotor-driven flow is developed to approximate the density transport dynamics.  
With this model, we show that the problem of controlled density transport can be posed as an optimal control problem which we solve using differential dynamic programming, an iterative trajectory optimization scheme. 
To apply this framework to the problem of steering a density using fixed rotors, we model the rotors as rotlets, the singularity solution of the Stokes equations associated with a point torque \cite{chwang1975hydromechanics,pozrikidis1992boundary}.  
This work is an extension of the authors' recent conference paper\cite{bt_cdc2023_densitycontrol}.  In this work, we seek to further highlight the fluid mechanical applications of the proposed method and use it to study the effects of boundaries on this fluid transport problem. Further, we show that the Perron-Frobenius operator associated with the time-varying flow field produced by the optimal control can be used to analyze the coherent sets and coherent structures of this flow. This enables us to make a useful qualititative connection between the optimal control and the dynamical structures which determine the Lagrangian transport, an area which has received increased interest in recent years \cite{krishna2022finite,krishna2023finite}.  

The rotlet singularity model has become commonly used as an approximation for flows generated by rotating bodies at small length and velocity scales.  Meleshko and Aref \cite{arefmeleshko} studied the flow generated by the so-called \emph{blinking rotlet} model consisting of two rotlets at fixed positions in a circular domain which run for a fixed time period in an alternating pattern, itself a Stokes flow alternative to the blinking vortex model introduced by Aref \cite{aref1984stirring}.  These models have received interest as minimalistic examples of the concept of chaotic advection, the notion that fluid particle trajectories in time dependent laminar flows can exhibit chaotic motions, even in two spatial dimensions. Van der Woude et al. \cite{van2007stokes} considered a similar blinking rotlet problem in a rectangular cavity and considered mixing by sinusoidal stirring patterns as well as the typical blinking pattern.  While these works introduce a time dependence by explicitly varying the strengths of fixed rotors in time, more recent works have studied effect of a time dependence in the fluid flow due non-stationary rotors, typically where each rotor is advected by the flow field generated by all other rotors \cite{lushi2015periodic,tallapragada2019rotlets,delmotte2019rollers}.  Recently, Piedra et al. \cite{piedra2023fluid} analyzed the flow produced by an electromagnetically driven rotor both experimentally and theoretically, also linking the mixing and transport properties of the flow to the associated coherent structures. In this work, while we only consider the case of fixed rotors, we develop methods to stir in a controlled way to steer a distribution to a desired location while minimizing the spread of the particle distribution.  

Several works have framed fluid mechanical transport problems as optimization or optimal control problems \cite{mathew2007optimal,dalessandro1999control,cortelezzi2008feasibility,lin2011optimal,hassanzadeh2014wall,zhang2020controlling}, with most of these focusing on optimizing mixing performance.  Mathew et al \cite{mathew2007optimal} studied the problem of optimally modulating (in time) a finite set of spatially varying force fields to optimize mixing over a fixed timespan and for a fixed action integral, using a conjugate gradient descent method to numerically approximate the optimal control. Zhang and Balasuriya\cite{zhang2020controlling} develop a method to determine an optimal spatiotemporally varying additive control velocity field for two problems: Lagrangian mixing and to drive trajectories to desired end states in a finite time. 
In this work, we present a numerical method to optimally modulate two flow fields (corresponding to rotors in fixed positions) in order to drive an initial distribution of fluid particles to a desired final distribution, as specified by the moments of the density function of the distribution. Further, we examine the structure of the optimal flow field by calculating the coherent sets and the associated flow structures produced by this flow field.  These results show that the optimal control typically produces a flow field which generates a transport barrier dividing the coherent sets which passes through the blob location at the initial time and connects to the target location at the final time, effectively directing the particle distribution toward the target. 

Our method relies on a finite-dimensional approximation of the Liouville operator, the infinitesimal generator of the semi-group of Perron-Frobenius operators \cite{lasota_1994}, which describe the density transport dynamics for a given flow map. 
The use of data-driven approximations of transfer operators in modelling fluid flows and in problems with actuation has been an active area of research in recent years \cite{ottorowley2021koopmancontrol,kaiser2020DataDrivenOperatorsReview}, with many of the most common methods having their origin in the analysis of fluid flows \cite{rowley2009spectral,mezic2013analysis}.
Refs. \onlinecite{froyland2016optimal,froyland2017optimalmixing} develop a convex optimization formulation based on transfer operators to determine optimal local perturbations of a flow field to enhance mixing of a fluid. 
Kl\"unker et al \cite{klunker2022open} recently studied mixing in open flows in terms of spectral properties of a finite-rank approximation of the Perron Frobenius operator. 
Sinha et al\cite{sinha2016operator} use the Perron Frobenius and Koopman generators associated with a given velocity field to choose an optimal location of release of a dispersant in the flow field. 
Brockett\cite{brockett2007optimal,brockett2012notes} proposed the optimal control of the Liouville equation with applications in ensemble control, but assumes a control input that can be varied arbitrarily in space and time.  Relatedly, Grover and Elamvazuthi \cite{grover2018optimalpert,grover2018_OTgenerators} use transfer operators and their generators in a graph-based approach to solving the optimal transport problem, motivated control problems for multi-agent and swarm systems, in which the control is also taken to vary spatiotemporally. The problem considered in this work can be viewed as a variation of those in \cite{grover2018optimalpert,grover2018_OTgenerators} with one significant distinction:  in this work the control input $u$ does not vary with the spatial location of the particle. The flow field is restricted to those that can be generated as linear combinations of the flow fields of two fixed micro-rotors and the strengths of the micro rotors in turn influence the flow field.

The remainder of the paper is structured as follows.  In Sec. \ref{sec:density_transport}, we review methods from the operator theoretic view of dynamical systems for modelling the transport of density functions through a dynamical system and present a numerical method for the computation of a finite dimensional approximation of the Liouville operator.  
In Sec. \ref{sec:gen_control} we demonstrate that this method can be naturally extended to account for the effects of actuation on a dynamical system, allowing the use of this framework to express the density transport problem as an optimal reference tracking problem.  
In Sec. \ref{sec:cohset_methods} we discuss how the operator theoretic methods relate to the computation of finite-time coherent sets for a time-varying flow field.
In Sec. \ref{sec:ddp} we briefly review the method of differential dynamic programming, an iterative trajectory optimization scheme which we implement to numerically solve this optimal control problem. 
In Sec. \ref{sec:freespace}, we implement these methods on the problem of steering a density of fluid particles using a pair of fixed microrotors.
In Secs. \ref{sec:wall} and \ref{sec:circle}, we study the effects of plane wall and circular boundaries on this transport problem in comparison to the case of an unbounded flow. In Sec. \ref{sec:two_blobs}, we consider the ability to manipulate multiple density functions simultaneously in this system. 

\section{Density Transport} \label{sec:density_transport}
In order to formulate the problem of controlling the motion of ensembles of fluid particles, we will first specify the distribution of such an ensemble by a density function.  In this section, we will review the methods used to study the evolution of such a density function over time, given that the individual particle motion is specified by a known dynamical system.

\subsection{Perron Frobenius operator and generator}
Consider a dynamical system 
\begin{equation}
    \frac{dx}{dt} = f(x) 
\end{equation}
on a measure space $(\XX, \A, \mu)$ where $x\in\XX$ is the state, $\XX\subset\RR^n$ is the state space, $\A$ is the Borel $\sigma$-algebra on $\XX$, and $\mu$ is a measure on $\XX$. Denote the time-$t$ flow map from an initial state $x_0$ by $\Phi^t(x_0)$.  We will further assume that the measure $\mu$ is absolutely continuous with respect to the Lebesgue measure, so that $\mu$ can be expressed in terms of a density, $\rho\in L_1(\XX)$, such that $d\mu(x) = \mu(dx) = \rho(x)dx$. 
With this, the Perron Frobenius operator, $\P^t:L_1(\XX)\mapsto L_1(\XX)$ corresponding to the flow $\Phi^t$ can be defined as the unique operator \cite{lasota_1994} such that 
\begin{equation}
    \int_A \P^t\rho(x) dx = \int_{(\Phi^t)^{-1}(A)} \rho(x) dx
\end{equation}
for any $A\in\A$, $t\geq 0$. The family of these operators, parameterized by time, $t$, have been shown to satisfy the properties of a semigroup \cite{lasota_1994}.  The infinitesimal generator of this semigroup, denoted here by $\L$, is known as the Liouville operator or the Perron-Frobenius generator, and defined as 
\begin{equation}\label{eq:generator_limit}
    \L\rho = \lim_{t\to 0} \frac{\P^t\rho - \rho}{t} = \lim_{t\to 0} \left(\frac{\P^t-\I}{t}\right) \rho
\end{equation}
where $\I$ is the identity operator.  Alternatively, as this operator expresses the deformation of a density function under an infinitesimal action of the operator $\P^t$, the Liouville operator can be thought of as expressing a continuity equation for the number of particles in the state space \cite{lasota_1994,cvitanovic_chaosbook}; that is,
\begin{equation}\label{eq:generator_continuity}
    \frac{\partial \rho}{\partial t} = \L \rho = -\nabla_x\cdot (\rho f)~.
\end{equation}
From this definition, we can immediately derive the following important property of the Liouville operator.
\begin{lemma} \label{lem_additivity}
Suppose the Liouville operator associated with a vector field $f_1:\XX \mapsto \RR^n$ is denoted by $\L_1$ and the Liouville operator associated with the vector field $f_2:\XX\mapsto \RR^n$ by $\L_2$, then the Liouville operator associated with the vector field $f(x) = f_1(x) + f_2(x)$, is $\L = \L_1 + \L_2$.
\end{lemma}

The numerical method used for the computation of the Perron-Frobenius operator and Liouville operator is derived from the relationship between the Perron-Frobenius operator and the Koopman operator.  The Koopman operator $\K^t:L^\infty(\XX) \mapsto L^\infty(\XX)$ is the operator which propagates observable functions $h\in L^\infty(\XX)$ forward in time along trajectories of the system and is defined as 
\begin{equation}
    \K^th = h\circ \Phi^t~.
\end{equation}
The Koopman and Perron-Frobenius operators are adjoint to one another, with the adjoint relationship given by 
\begin{equation} \label{eq:adjoint}
    \int_{\XX}[\K^th](x)\rho(x)dx = \int_{\XX}h(x)[\P^t\rho](x) dx \,.
\end{equation}

\subsection{Numerical approximation} \label{sec:PF_edmd}
One of the most common methods of approximating the Perron-Frobenius operator is a set-oriented approach known as Ulam's method\cite{ulam1960collection}, in which a domain of interest is discretized into cells, a large number of short-time trajectories are simulated, and then the operator is computed as the matrix containing the approximate transition probabilities between the cells\cite{dellnitz2001GAIOalgorithms}.  It has been shown that this method can be viewed as a Galerkin projection of the Perron Frobenius operator onto the function space spanned by indicator functions corresponding to the discrete cells \cite{Klus16_onthenumerical}. 
In recent works involving numerical approximation of the Koopman operator, one of the most common approaches is that of extended dynamic mode decomposition (EDMD) \cite{williams2015data}, in which the operator is computed by solving a least squares problem, which can also be viewed as a Galerkin projection of the operator onto a function space spanned by a predefined set of basis functions \cite{williams2015data,Klus16_onthenumerical}.  
By exploiting the adjoint relationship between the Perron Frobenius and Koopman operators, it has been shown that methods typically used for one operator can be used to compute the other.  Based on this idea, recent works have developed variations of EDMD for the computation of the Perron-Frobenius operator \cite{vaidya_nsdmd,goswami2018constrained,Klus16_onthenumerical}.  
In this work, we also implement EDMD for the computation of the Perron-Frobenius operator, which we outline below, largely following Klus \emph{et al}. \cite{Klus16_onthenumerical}.  

The method requires a predefined dictionary $\DD$ of $k$ scalar-valued basis functions, $\DD = \{\psi_1, \psi_2, \dots, \psi_k\}$, where $\psi_i:\XX\mapsto\RR$ for $i = 1, \dots, k$ and trajectory data collected from the dynamical system with fixed timestep, $\Delta t$, arranged into snapshot matrices as 
\begin{align}
X &= 
\begin{bmatrix} \label{eq:snapshotx}
x_1 & , ~\cdots~ , & x_m 
\end{bmatrix}\\
Y & = 
\begin{bmatrix} 
x_1^+ &, ~\cdots~, & x_m^+ 
\end{bmatrix}
\end{align}
where the subscript $i=1,\dots, m$ is a measurement index and $x_i^+ = \Phi^{\Delta t}(x_i)$. 

Then, given an the observable function $h$ and density $\rho$, these functions are approximated by their projections onto the space spanned by elements of $\DD$ as
\begin{align}
    h(x) &\approx \hat{h}^T\Psi(x)\\
     \rho(x) &\approx \Psi^T(x) \hat{\rho} \label{eq:dens_proj} 
\end{align}
where $\hat{h},\,\hat{\rho}\in\RR^k$ are column vectors containing the projection coefficients and $\Psi:\XX\mapsto\RR^k$ is a column-vector valued function where the elements are given by $[\Psi(x)]_i = \psi_i(x)$. 
Substituting these expansions into Eq. \eqref{eq:adjoint} yields
\begin{equation}
    \int_{\XX} \K^{\Delta t}[\hat{h}^T\Psi]\Psi^T\hat{\rho} \, dx 
    = 
    \int_{\XX} \hat{h}^T\Psi \P^{\Delta t}[\Psi^T\hat{\rho}]\,dx\,.
    \label{eq:adj_expanded}
\end{equation}
Then noting that $[\K^{\Delta t}\Psi](x) = \Psi(x^+)$ and assuming that $\P^{\Delta t}$ can be approximated by a matrix $P$ operating on the coordinates $\hat{\rho}$, it is clear that in the limit of a large dataset $m\to \infty$, the above expression becomes 
\begin{equation}
    \Psi_Y\Psi_X^T = \Psi_X\Psi_X^TP + e
\end{equation}
where $e$ is a residual error arising due to the matrix approximation of $\P^{\Delta t}$ by $P$.  
This can be posed as a least-squares problem for the matrix $P$
\begin{equation}
    \min_P \|\Psi_Y\Psi_X^T - \Psi_X\Psi_X^TP\|_2^2
\end{equation}
where $\Psi_X$,$\Psi_Y \in\RR^{k\times m}$ are matrices with columns containing $\Psi$ evaluated on the columns of $X$ and $Y$ respectively.  The analytical solution of this least squares problem is 
\begin{equation} \label{eq:pf_lsq}
    P = \left(\Psi_X\Psi_X^T\right)^\dagger\Psi_Y\Psi_X^T 
\end{equation}
where $(\cdot)^\dagger$ is the Moore-Penrose pseudoinverse. 

Given this matrix approximation of the operator, $P$, if the timestep $\Delta t$ chosen in the data collection is sufficiently small, the corresponding matrix approximation $L$ of the Liouville operator can be approximated based on the limit definition of the generator in Eq. \ref{eq:generator_limit}. as 
\begin{equation}\label{eq:gen_approx}
    L\approx \frac{P - I_k}{\Delta t}
\end{equation}
where $I_k$ is the $k\times k$ identity matrix.  The matrix approximation $P$ of the operator $\P^{\Delta t}$ approximates the propagation of a density function $\rho$ by advancing the projection coordinates $\hat{\rho}$ forward for a finite time, $\Delta t$.  Similarly, the matrix approximation $L$ of the generator $\L$ approximates the infinitesimal action of the operator $\P^t$ by approximating the time derivative of the projection coordinates 
\begin{equation}
    \frac{d\hat{\rho}}{dt} = L\hat{\rho}\,.
\end{equation}

\subsection{Extension to controlled systems} \label{sec:gen_control} 
In the field of control theory, much attention has been given in recent years to applications the Koopman operator to control systems \cite{korda2018linear,kaiser2020DataDrivenOperatorsReview, ottorowley2021koopmancontrol}, including several recent works which have noted the usefulness of formulating the problem in terms of the Koopman generator, rather than the Koopman operator \cite{goswami2017globalbilin,klus2020gedmd,rowley2020interpolated,bruder2021advantages,folkestad2021koopmanNMPC}.  Such a formulation in terms of the Koopman generator typically results in a lifted system that is bilinear in the control and lifted state, as the effect of the control vector fields is expressed in a way that is also dependent on the lifted state. This approach allows for a better approximation of the effects of control as compared to other common approaches\cite{bruder2021advantages}, especially for systems in control-affine form 
\begin{equation}\label{eq:controlaffinesys}
    \frac{dx}{dt} = f(x) + \sum_{i = 1}^{n_c}g_i(x)u_i
\end{equation}
where the $u_i$ are control inputs and $n_c$ is the number of control inputs affecting the system.
Here we apply a similar approach to the density transport problem, expressed in terms of the Perron-Frobenius generator. As shown by Peitz et al. \cite{rowley2020interpolated} for the Koopman generator, by the property of the Perron-Frobenius generator given in Lemma \ref{lem_additivity}, if the dynamics are control-affine, then the generators are also control affine, as can be seen by application of Eq. \ref{eq:generator_continuity}.  This leads to density transport dynamics of the following form
\begin{equation}\label{eq:dens_dyn}
    \frac{\partial \rho}{\partial t} = \L_0\rho + \sum_{i=1}^{n_c}u_i\B_i\rho
\end{equation}
where $\L_0$ is the Perron Frobenius generator associated with the vector field $f(x)$ and similarly, the $\B_i$ are the Perron Frobenius generators associated with the control vector fields $g_i(x)$. 
Therefore, given the finite dimensional approximation of these generators, we can approximate the density transport dynamics as 
\begin{equation} \label{eq:dens_dyn_fin}
    \frac{d\hat{\rho}}{dt} = L_0\hat{\rho} + \sum_{i=1}^{n_c} u_iB_i\hat{\rho}
\end{equation}
where the matrices $L_0$ and $B_i$ are the matrix approximations of the operators in Eq. \ref{eq:dens_dyn}.  These matrix approximations can be computed using the method outlined in Sec. \ref{sec:PF_edmd} for uncontrolled systems. This is done by first computing $L_0$ by Eq. \ref{eq:gen_approx} using trajectory data from the system with all control inputs set to zero.  Once $L_0$ is found, each of the $B_i$ can be computed similarly by first computing a matrix $L_i$ by Eq. \ref{eq:gen_approx} using trajectory data from the system collected with $u_i = 1$ and $u_j = 0$ for $j\neq i$. This matrix $L_i$ approximates the Liouville operator corresponding to a vector field $f + g_i$.  The matrix approximation $B_i$ of the operator corresponding to the vector field $g_i$ alone is then found using Lemma \ref{lem_additivity} as $B_i = L_i - L_0$. 

For the systems of microrotors considered in this work, the control inputs are taken to be the strengths $\gamma_i$ of a pair of micro-rotors and the states are taken to be the position coordinates of a fluid particle.  In this application, it will be shown (see, e.g. Eq. \ref{eq:rotlet_vel}), that the control system is drift-free.  That is, the vector field $f = 0$ in Eq. \ref{eq:controlaffinesys}, and therefore, the corresponding Liouville operator $\L_0 = 0$, as well for these systems. 
This is due to the typical quasistationary assumption of Stokes flows, 
which indicates that any change in the flow field is established instantaneously, without transience \cite{happelbrenner}. 
\subsubsection{Propagation of moments}
In what follows, the problem of driving an initial density to a desired final density will be posed as an optimal control problem.  The control inputs for this problem are the strengths of a finite number of micro-rotors, meaning that this problem involves steering a function using only a finite number of control inputs.  To make this problem more tractable, we instead consider the problem of steering the moments of the density function to match the moments of a desired final density function.  In the remainder of this section, an approximation of the moments of a density function $\rho(x)$ are derived in terms of the projection of $\rho$ onto the space spanned by the elements of $\DD$. 

Given a projection of $\rho$ onto $\DD$, as in Eq. \ref{eq:dens_proj}, the first moment (mean), $m_1$ is written as 
\begin{equation} \label{eq:mean_int}
    m_1^i = \int x^i\rho(x) dx = \hat{\rho}^T\int x^i\Psi(x) dx
\end{equation}
where we use the superscript $i$ in the moment to indicate the coordinate index and the subscript indicates the order of the moment being considered.  Therefore, the first moment of $\rho$ can be approximated as a linear combination of the means of the dictionary functions in $\Psi$, weighted by the projection coefficients $\hat{\rho}$. 
This is also true for higher order raw moments, whereas higher order central moments become polynomial in $\hat{\rho}$ due to their dependence on the mean. 
Since $\hat{\rho}$ will be treated as the `lifted state' in the control formulation, it is desirable to consider moments which are linear in $\hat{\rho}$, so for this reason we will work with raw moments in what follows.  

Here, for the dictionary functions, we use Gaussian radial basis functions of the form 
\begin{equation}\label{eq:gaussrbf}
    \psi_l(x) = \exp\left(-\frac{(x-c_l)^T(x-c_l)}{2s^2}\right)
\end{equation}
where $c_l$ is the center of the $l^\text{th}$ basis function, and $s$ is a scaling parameter affecting the spread.  Computing the integral in Eq. \ref{eq:mean_int}, in terms of this dictionary, the mean is approximated as 
\begin{equation}\label{eq:m1}
    m_1^i = 2\pi s^2\sum_{l=1}^k\hat{\rho}_lc^i_l
\end{equation}
where $c_l^i$ is the $i^\text{th}$ coordinate of the $l^\text{th}$ basis function center.
Similarly, the second raw moment can be written as 
\begin{equation}\label{eq:m2}
m_2^{ij} = \int x^ix^j\rho(x)dx = \hat{\rho}^T\int x^ix^j\Psi(x)dx
\end{equation}
where the last integral reduces to 
\[
\int x^ix^j\psi_l(x)dx = 
\begin{cases}
2\pi s^2(s^2 + (c_l^i)^2) & i=j\\[1ex]
2\pi s^2c_l^ic_l^j & i\neq j
\end{cases}
\]
for a given basis function $\psi_l(x)$ where superscripts $i$ and $j$ are coordinate indices. 

\subsection{Finite-time coherent set detection}  \label{sec:cohset_methods}
For autonomous dynamical systems, methods based on the Perron-Frobenius operator have been used to compute invariant or almost invariant sets of the system \cite{dellnitz1999approximation}.  This is typically done by studying eigenfunctions of the Perron-Frobenius operator with eigenvalues, $\lambda\approx 1$.  Such eigenfunctions correspond to invariant or almost invariant densities, which describe groups of states which are left nearly unchanged by the flow of the system.
These methods have also been extended to time-varying systems, in which the goal is to identify finite-time coherent sets\cite{froyland2010transport,williams2015identifying,allshouse2015lagrangian, tallapragada_cnsns_2013}.  Such sets are defined as sets in the state-space which are maximally coherent, or minimally dispersive, over a certain finite time interval. That is, they describe sets of states which may be transported as a whole by the flow, but with minimal transport outside of the coherent set or between coherent sets.  These methods are also closely related to the Perron-Frobenius operator and are commonly seen as a probabilistic alternative to geometric methods related to the identification of invariant manifolds, dominant material lines, or Lagrangian coherent structures (see Refs. \onlinecite{allshouse2015lagrangian,hadjighasem2017critical} for a review). 

Here, we will apply the methods of Ref. \onlinecite{williams2015identifying} to the time-varying flow field generated by the solution to the optimal control problem to elucidate the flow structures associated with the optimal control.  In this section, we will briefly summarize the method for the detection of coherent structures used here and its relation to the finite-dimensional operator approximation defined in the previous section. 

We assume that a dataset is given of $m$ points, $\{(x_i,y_i)\}_{i=1}^m$, where $x_i$ is the position of the $i^{\text{th}}$ particle at the initial time, $t_0$ and $y_i$ is the position of the particle at a later time $t_f$. That is, $y_i = \Phi_{t_0}^{t_f}(x_i)$, where $\Phi_{t_0}^{t_f}$ is the flow map associated with the non-autonomous system from time $t_0$ to $t_f$. Given that the data lies in a set $X$ at time $t_0$ and a set $Y$ at time $t_f$, our goal is to partition this dataset into two sets, $X_1$ and $X_2$ at time $t_0$ and $Y_1$ and $Y_2$ at time $t_f$, such that points in $X_1$ are mapped into $Y_1$ by the flow and points in $X_2$ are mapped into $Y_2$. This partition is designed by constructing partition functions $f_X$ and $f_Y$ which partition the space based on their sign. For example, we can define $X_1 = \{x\in X|f_X(x) > 0\}$. 
Then the problem of identifying coherent sets can be framed as choosing the functions $f_X$ and $f_Y$ to maximize the objective 
\begin{equation}  \label{eq:g_cohobj}
    g(f_X,f_Y) = \frac{1}{m}\sum_{i = 1}^m f_X(x_i)f_Y(y_i)
\end{equation}
which can be thought of as an approximation of the an inner product 
\begin{subequations}
    \begin{align}
        g(f_X,f_Y) &\approx \langle f_X,\K_{t_0}^{t_f}f_Y \rangle = \int_X f_X(x) f_Y\left(\Phi_{t_0}^{t_f}(x)\right)dx \\
&\approx \langle \P_{t_0}^{t_f}f_X,f_Y\rangle= \int_Y f_X\left(\Phi_{t_f}^{t_0}(y)\right)f_Y(y)dy
    \end{align}
\end{subequations}
where $\K_{t_0}^{t_f}$ and $\P_{t_0}^{t_f}$ are the Koopman and Perron-Frobenius operators associated with this time-varying flow and $\Phi_{t_f}^{t_0} = \left(\Phi_{t_0}^{t_f}\right)^{-1}$.
Note that this objective is only reasonable if an overall scale is imposed on the magnitude of the functions $f_X$ and $f_Y$. 
If we approximate the partition functions $f_X$ and $f_Y$ by their projection onto the space spanned by the dictionary $\DD$,
\[
f_X(x) \approx \Psi^T(x)a , \qquad \qquad 
f_Y(y) \approx \Psi^T(y)\at
\]
then the objective is approximated as 
\begin{equation}
    g(f_X,f_Y) \approx \frac{1}{m}\sum_{i = 1}^m a^T\Psi(x_i)\Psi^T(y_i)\at = a^TA\at
\end{equation}
where $A = \frac{1}{m}\Psi_X\Psi_Y^T$.  If we impose a scale by requiring that $a^Ta = \at^T\at = 1$, then this maximization can be solved by singular value decomposition, with the optimal $a$ and $\at$ given by left and right singular vectors, respectively, as shown in Refs. \onlinecite{froyland2010transport,williams2015identifying}. 
This problem can be solved trivially by choosing $f_X$ to be uniform over $X$ and choosing $f_Y$ to be uniform over $Y$ -- this solution typically corresponds to the singular vector associated 
with the largest singular value. Therefore, the singular vectors associated with the 2nd largest singular value give the optimal non-trivial solution, which divides the domain into partitions of roughly equal size \cite{williams2015identifying}.

\section{Control formulation } \label{sec:ddp}
In Sec. \ref{sec:gen_control}, it was shown that the problem of steering a density $\rho$ to a desired final density can be expressed as an output tracking problem on a lifted, bilinear system given by Eq. \ref{eq:dens_dyn_fin}, where the projection coefficients $\hat{\rho}$ can be interpreted as a lifted state. 
Then, if the first and second raw moments are taken to be the relevant output, 
\begin{equation} \label{eq:momentvec}
    y = \begin{bmatrix}
        m_1^1 & m_1^2 & m_2^{11}& m_2^{22}& m_2^{12}
    \end{bmatrix}
    ^{T}
\end{equation}
this can be expressed linearly in the lifted state, $y = C\hat{\rho}$, where the elements of the output matrix $C$ are given by rewriting Eqs. \ref{eq:m1}, \ref{eq:m2} in matrix form. 

For the optimal output tracking problem, we consider a discrete time optimal control problem 
\begin{subequations}
\begin{align} \label{eq:optimization}
    \min_{u_1,u_2,\dots,u_{H-1}} &\sum_{t=1}^{H-1}l(\rh_t,u_t) + l_H(\rh_H)\\[1ex]
    \mathrm{s.t.} \qquad & \hat{\rho}_{t+1} = F(\hat{\rho}_t,u_t) \label{eq:disc_dyn} \\
    & y_t = C\rh_t
\end{align}
\end{subequations}
where $H$ is the number of timesteps in the time horizon and Eq. \ref{eq:disc_dyn} represents the discrete time version of Eq. \ref{eq:dens_dyn_fin}.

In particular, for output tracking, we consider in-horizon and terminal cost functions $l$ and $l_H$ of the following quadratic forms
\begin{subequations}\label{eq:cost_both}
    \begin{align}  
        l(\rh_t,u_t) &= (y_t - y^{\mathrm{ref}}_{t})^TS(y_t - y^{\mathrm{ref}}_{t}) + u_t^TRu_t 
        \label{eq:cost}
        \\
        l_H(\rh_H) &= (y_H - y^{\mathrm{ref}}_{H})^TS_H(y_H - y^{\mathrm{ref}}_{H}) \label{eq:termcost}
    \end{align}
\end{subequations}
where $S$, $R$, and $S_H$ are weighting matrices which define the penalty weight on tracking error, control effort, and error in the terminal state, respectively. Since the output $y$ is linear in the lifted state $\rh_t$, this cost can be rewritten as a quadratic cost in terms of $\rh_t$, with an added linear term.  

It is well known that for optimal control problems on bilinear systems with quadratic cost, an effective way of solving the problem is by iteratively linearizing and solving a finite time linear quadratic regulator (LQR) problem about a nominal trajectory, utilizing the Ricatti formulation of that problem \cite{hofer1988iterative}.  For this reason, we solve the optimal control problem using differential dynamic programming (DDP) \cite{tassa2012synthesis,yakowitz1984computational}, which is closely related to the method of iterative LQR.  We briefly recount the primary steps of this algorithm below.  

\begin{figure*}
    \centering 
    \includegraphics[width=\linewidth]{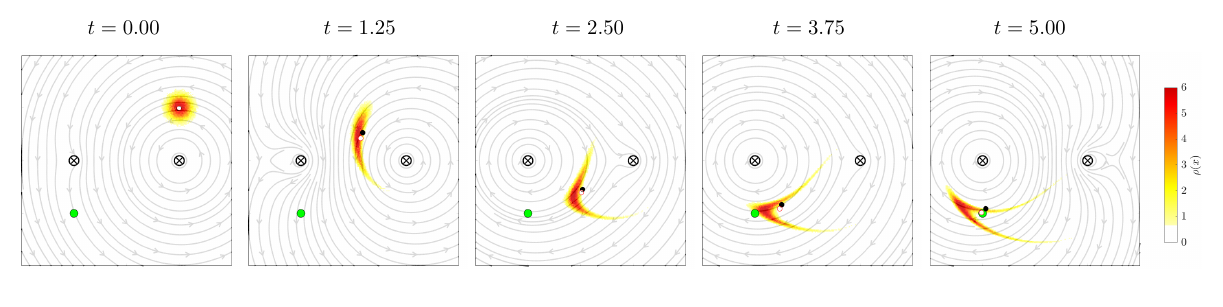}
    \vspace{-1cm}
    \caption{Controlled transport of a distribution of fluid particles by two micro-rotors fixed at $(-1,0)$ and $(1,0)$ from an initial density $\rho(x(0)) = \N((1,1),0.025I_2)$ to a target mean at $(-1,-1)$ (green circle). White filled circle indicates the sample mean and black filled circle indicates the mean predicted by the proposed method. Streamlines depict the flow field produced by the rotor control at the instant shown. }
    \label{fig:freespace_stream}
\end{figure*}

DDP computes a locally optimal control around a nominal trajectory by minimizing a quadratic approximation of the value function along this trajectory, and then doing this iteratively about the new trajectories obtained by applying the locally optimal control.  First define the value function $V(\rh_t,t)$ at time $t$ as,
\begin{equation}
V(\rh_t,t) = \min_{u_t} [ l(\rh_t,u_t) + V(\rh_{t+1},t+1)]   
\end{equation}
which expresses the optimal cost-to-go from $\rh_t$, where $V(\rh_H,H) = l_f(\rh_H)$.
Denote by $Q(\delta\rh, \delta u)$ the change in the value function due to applying change in control input $\delta u$ about the nominal trajectory and consider its quadratic approximation 
\begin{equation}
\begin{split}\label{eq:quadQ}
    Q(\delta\rh, \delta u) \approx &
    \:Q_{\rh}\delta\rh + Q_u^T\delta u + \delta \rh^T Q_{\rh u}\delta u \\[1ex]& + \frac{1}{2}\delta\rh^TQ_{\rh\rh}\delta \rh + \frac{1}{2}\delta u^T Q_{uu}\delta u 
\end{split}
\end{equation}
where these derivatives are given by 
\[
\begin{split}
Q_{\rh} &= l_{\rh}+ F_{\rh}^T V_{\rh}'\\
Q_{u} &= l_{u}+ F_{u}^T V_{\rh}'\\
Q_{\rh\rh} &= l_{\rh\rh}+ F_{\rh}^T V_{\rh \rh}'F_{\rh} + V_{\rh}'\cdot F_{\rh\rh}\\
Q_{uu} &= l_{uu}+ F_{u}^T V_{\rh\rh}'F_{u} + V_{\rh}'\cdot F_{uu}\\
Q_{\rh u} &= l_{\rh u}+ F_{\rh}^T V_{\rh \rh}'F_{u} + V_{\rh}'\cdot F_{\rh u}\\
\end{split}
\]
where the notation $(\cdot)'$ indicates the next time step.
The algorithm proceeds by computing these derivatives by recursing backward in time along the nominal trajectory from the end of the horizon.
At each iteration, the control policy is improved by optimizing this quadratic expansion with respect to $\delta u$
\begin{equation}
    \delta u^* = \arg \min_{\delta u}Q(\delta \rh,\delta u) = -Q_{uu}^{-1}\left(Q_u + Q_{u \rh}\delta \rh\right)
\end{equation}
This can be seen as providing a descent direction in the space of control policies.  An updated nominal control is then computed by a line search over a stepsize parameter $\alpha$ to update the policy, that is 
\[
u_{\text{new}} = u - \alpha Q_{uu}^{-1}Q_u - Q_{uu}^{-1}Q_{u \rh}\delta \rh
\] 
and this new control is applied to obtain a new nominal trajectory, and this procedure is iterated until the relative change in cost falls to less than a specified tolerance. For full details of the algorithm, the reader should refer to Refs. \cite{tassa2012synthesis,yakowitz1984computational}.

\begin{figure*}
    \centering
    \includegraphics[width = 0.75\linewidth]{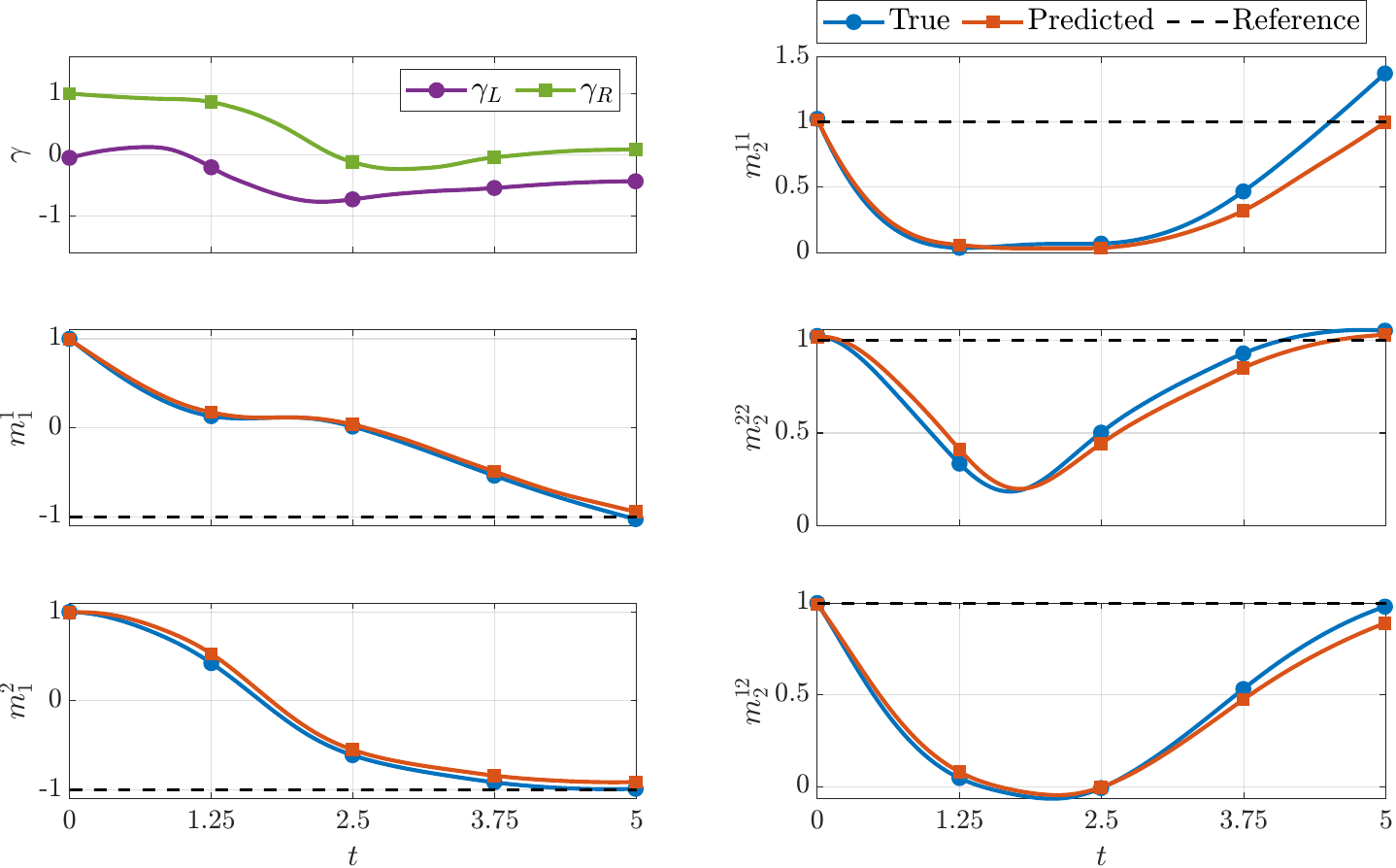}
    \caption{Rotor controls and fluid particle distribution moments for transport by rotors in free space as shown in Fig. \ref{fig:freespace_stream}. Left, Top: Rotor strengths for the left rotor ($\gamma_L$) and right rotor ($\gamma_R$). Left, Bottom: Mean of particle distribution from sample shown in Fig. \ref{fig:freespace_stream} (`True') and as predicted using the Liouville operator (`Predicted').  Right: Second raw moment. Markers correspond to instants shown in Fig. \ref{fig:freespace_stream}. }
    \label{fig:freespace_umu}
\end{figure*}

\section{Transport by rotors in free space} \label{sec:freespace}

To describe the fluid flow produced by a microscale rotor, we employ a model of a point torque in a two dimensional Stokes flow.  Mathematically, this flow is described by a rotlet \cite{pozrikidis1992boundary}, whose stream function is given by 
\begin{equation}
    \psi(\x) = -\gamma \log |\x - \x_r|
\end{equation}
where $\x = (x,y)$ is the position a point in the fluid, $\x_r = (x_r,y_r)$ is the position of the rotlet, and $\gamma$ is the strength of the rotlet.  Physically, $\gamma$ describes the magnitude of the point torque or the angular velocity of the rotor. The linearity of Stokes flows allows for the velocity fields produced by multiple rotlets to be determined by superposition of the velocity field produced by each rotlet individually.  Therefore, for $n_r$ rotors, the resulting fluid flow  results in the following fluid velocity field: 
\begin{equation} \label{eq:rotlet_vel}
    \mathbf{u}(\x) = -\sum_{i=1}^{n_r} \left(\gamma_i\hat{k} \times \frac{~\x - \x_i~}{r_i^2}\right)
\end{equation}
where $\x_i$ is the location of the $i$-th rotlet and $r_i = |x - x_i|$.  Clearly, this results in a flow with a singularity at $\x_r$, circular streamlines around the singularity with counterclockwise flow for positive $\gamma$, and a fluid velocity that decays as $r^{-2}$ going away from the rotor. 

Here, we consider the case of rotors fixed in place on the $x$-axis at $(-1,0)$ and $(1,0)$, respectively, denoting their strengths by $\gamma_L$ and $\gamma_R$ for left and right.  We consider a problem of manipulating a collection of fluid particles initially distributed at time $t=0$ according to a normal distribution, with mean $m_1(0) = (1,1)$ and covariance $\Sigma(0) = 0.025I_2$, where $I_2$ is the $2\times2$ identity matrix. That is, $\rho(x(0)) = \N\big((1,1),0.025\,I_2\big)$.  From this initial fluid particle distribution, we seek a sequence of rotor strengths over a timespan of 5 time units to drive the fluid particles to a final distribution with a mean of $m_1(5) = (-1,-1)$, while minimizing the variance.  For this, we use the relationship between the second raw moment and the variance
\begin{equation}
    m_2^{ij} = \sigma^{ij} + m_1^im_1^j
\end{equation}
where the $\sigma^{ij}$ are the elements of the covariance matrix $\Sigma$, to convert the desired final variance to a desired second moment. 

With this, a rotor control sequence is found by solving the optimization problem as in Eq. \ref{eq:optimization} using the DDP scheme described in Sec. \ref{sec:ddp}.  In solving this, the cost function weights are chosen to be $S = 0.1I_{5}$, $R = I_2$, and $S_H = 10^3I_5$. That is, the error in the moments is penalized very little from $t = 0$ until $t=5$, with a large penalty placed on the moments at $t=5$.  This choice allows the optimizer the flexibility to steer the distribution in a way that may temporarily increase the error if it results in a lower error in the moments at $t = 5$. 

For the computation of the Liouville operators for this case, data is collected by simulating a grid of 2500 initial conditions, evenly spaced over $[-2,2]^2$ forward for time interval $\Delta t = 0.005$.  The Perron-Frobenius operators are computed using Eq. \ref{eq:pf_lsq} with this trajectory data and a $25\times25$ grid of Gaussian radial basis functions with centers evenly spaced over the same domain, excluding small radii around the rotors.  The Liouville operators are obtained from this using Eq. \ref{eq:gen_approx} as described in Sec. \ref{sec:gen_control}.

Fig. \ref{fig:freespace_stream} shows the effect of the rotor control on the motion of a distribution of $10^4$ fluid particles sampled according to the initial density and displayed as a histogram approximation of the density. 
The rotor positions are indicated by the circle-cross and the position of the target mean is shown by the green circle.  The white-filled circle indicates the mean of this sample, while the black-filled circle indicates the mean as predicted using the Liouville operator.  The streamlines in the figure indicate direction of the fluid velocity field produced by the rotors at the indicated time instant.  Fig. \ref{fig:freespace_umu} shows the rotor strengths $\gamma_L$ and $\gamma_R$ selected by the DDP algorithm. For the first 1.25 seconds, the rightmost rotor has a positive strength of near $1$ to generate a counterclockwise flow, pulling the distribution of particles toward the origin, while the leftmost rotor has a low strength near zero. As the distribution nears the origin, the strength of the right rotor decreases, while the magnitude of the strength of the left rotor increases to generate a clockwise flow, which pulls the distribution toward the target mean. 

Also shown in Fig. \ref{fig:freespace_umu} are plots of the elements of the first and second moment over time as computed from the sample shown in Fig. \ref{fig:freespace_stream} (labelled `True') and as predicted using the finite approximation of the Liouville operators (labelled `Predicted').

\begin{figure*}
        \centering
   \includegraphics[width=\linewidth]{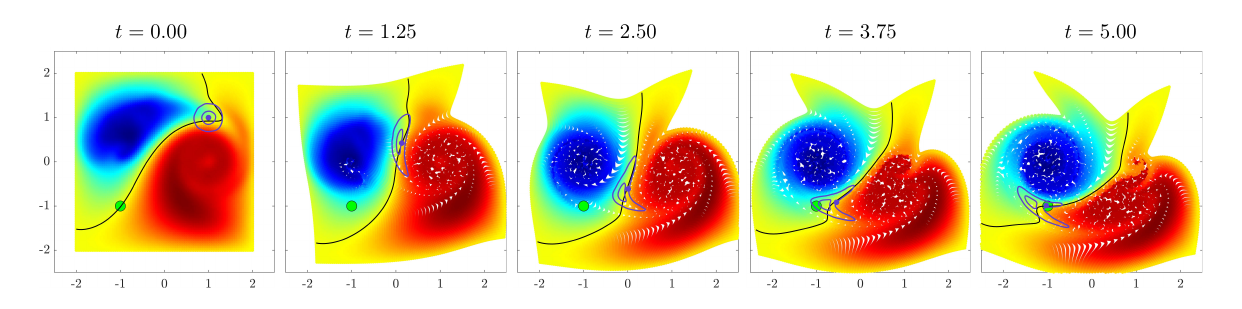}
     \vspace{-0.7cm}
    \caption{Finite-time coherent sets produced by the optimal rotor-driven flow field corresponding to the case shown in Figs. \ref{fig:freespace_stream} and \ref{fig:freespace_umu}.  The sequence shows the time evolution of $10^4$ particles initially placed on a uniform grid on $[-2,2]^2$ and colored according to the partition function $f_X$, as approximated by the left singular vector of $A=\frac{1}{m}\Psi_X\Psi_Y^T$.  The black line in the first image depicts the $f_X=0$ level set, and the sequence shows the evolution of this line as the dataset is advected by the flow. The purple marker and lines show the mean value and level sets of the density function corresponding to values of the initial density at one and two standard deviations from the mean, respectively.  }
    \label{fig:cohsets_evolution_x1y1}
\end{figure*}

\subsection{Finite time coherent sets} 
With the optimal control determined, Eq. \ref{eq:rotlet_vel} gives a nonautonomous dynamical system.  We can then apply the methods outlined in Sec. \ref{sec:cohset_methods} to this system  to identify coherent sets to better understand the underlying structure of the flow field produced by the optimal control. For this computation, we use a dataset of $10^4$ data pairs, initially spaced on a uniform grid over $[-2,2]^2$.  For the basis, we use a set of 2501 basis functions consisting of Gaussian radial basis functions uniformly spaced on a $50\times50$ grid over $[-2,2]^2$ and the constant function, $\psi = 1$.  
Fig. \ref{fig:cohsets_evolution_x1y1} shows the time evolution of the data set, with the points colored according to the partition function $f_X$ as approximated by the 2nd left singular vector of $A = \frac{1}{m}\Psi_X\Psi_Y^T$.  Also shown is the evolution of the $f_X=0$ contour, which approximates the barrier between the coherent sets, as depicted by the black line.  Finally, level sets and mean of the density function, $\rho(x(t))$, as approximated from a sample of $10^4$ points from the initial density, are shown by the purple contours and purple markers, respectively.  The level sets shown correspond to values of the initial density at one and two standard deviations from the mean, respectively. 

To quantify the coherence of the sets identified by the partition functions $f_X$ and $f_Y$, we use a modification of the objective in Eq. \ref{eq:g_cohobj}, which only considers the sign of $f_X$ and $f_Y$, 
\begin{equation}\label{eq:mod_g}
    \bar{g}(f_X,f_Y) = \frac{1}{m}\sum_{i = 1}^m \sign(f_X(x_i))\sign(f_Y(y_i))
\end{equation}
which effectively gives the fraction of the data points which are classified correctly by the partition functions (for which the partition functions do not change sign from initial to final time).  For the case shown in Fig. \ref{fig:cohsets_evolution_x1y1}, we have $\bar{g} = 0.9904$. 

\begin{figure}
    \centering
    \includegraphics[width=0.49\linewidth]{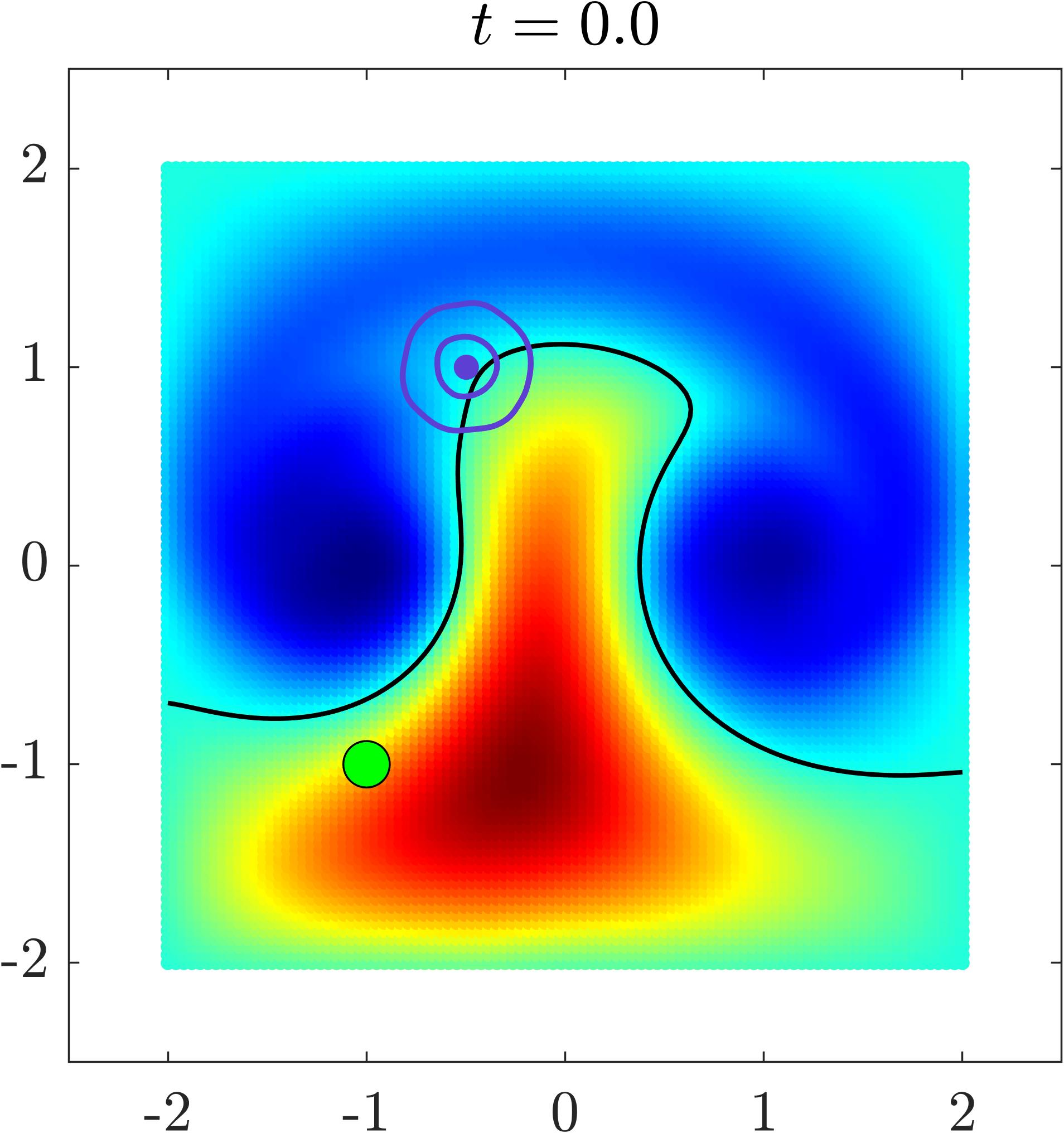}
    \includegraphics[width=0.49\linewidth]{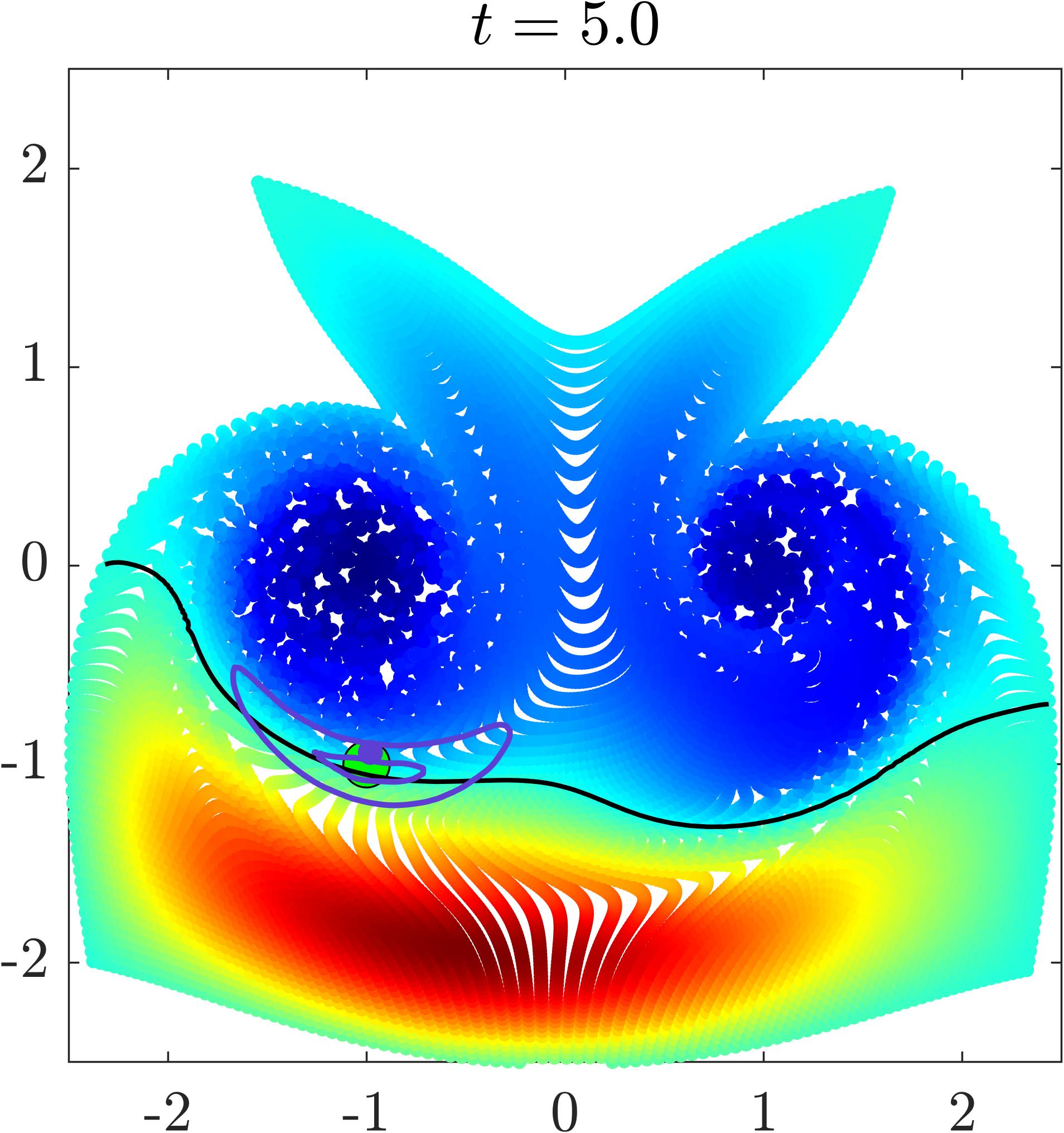}
    \caption{Finite-time coherent sets produced by the optimal rotor-driven flow field with initial density $\rho(x(0)) = \N((-0.5,1),0.025I_2)$ .  The subfigures show the $10^4$ data points at the initial time $t=0$ and final time $t=5$, colored according to the 3rd left singular vector of $A$, along with the $f_X = 0$ contour (black line).  Purple contours depict the mean value and level sets of the density function corresponding to values of the initial density at one and two standard deviations from the mean, respectively.}
    \label{fig:cohsets_xm05y1}
    \vspace{-1em}
\end{figure}
This computation of the coherent sets shows that the flow field generated by the optimal control is such that a transport barrier is formed over the 5s time interval, with the barrier passing through the particle distribution at the initial time and connecting to the target location at the final time. 
Many previous works \cite{inanc2005optimal,senatore2008fuel,ramos2018lagrangian,krishna2022finite} have studied the relationship between the optimal control problem of steering a particle efficiently in an unsteady flow and the coherent structures associated with that flow.  Typically in these studies, the problem being considered is motivated by the efficient navigation of an underwater vehicle to a target in an unsteady ocean flow.
For this reason, the control input is usually taken to be a propulsive velocity which is added to the unsteady flow field, as could be generated by a thruster onboard an underwater vehicle, and coherent structures associated with the unsteady flow are used to identify efficient routes.  
Our work takes a different perspective, where instead of controlling individual particles in a given unsteady flow field, we solve an optimal control problem to determine the optimal time-varying  flow field to steer the initial particle distribution to the target, where the unsteady flow field is constrained to be a superposition of flow fields produced by the two rotors at each time instant.  In previous works \cite{inanc2005optimal,krishna2022finite}, it was seen that the optimal routes of an underwater vehicle tend to follow the coherent structures which guide the particle towards the target for energy-optimal navigation.  Here we see that the optimal flow field produces a flow structure which guides the distribution of particles from the initial condition to the target, as shown in Fig. \ref{fig:cohsets_evolution_x1y1}.  

\begin{figure*}
    \centering
    \includegraphics[width=\linewidth]{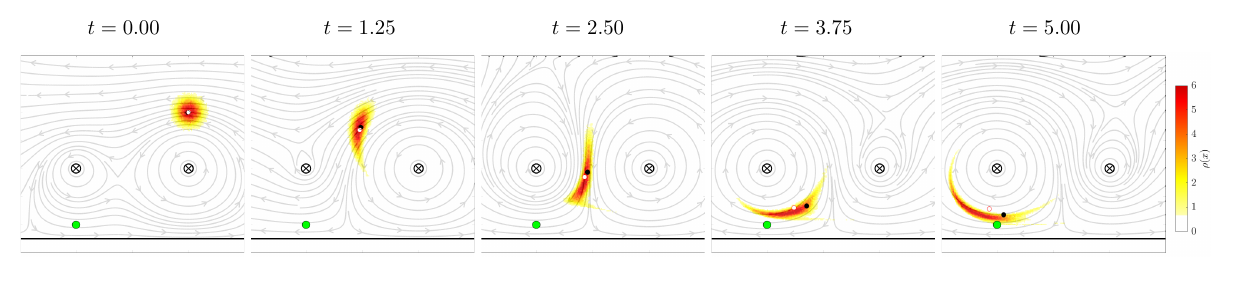}
    \vspace{-1cm}
    \caption{Controlled transport of a distribution of fluid particles near an infinite plane wall at $y = -1.25$ (black line) by two micro-rotors fixed at $(-1,0)$ and $(1,0)$  from an initial density $\rho(x(0)) = \N((1,1),0.025I_2)$ to a target mean at $(-1,-1)$ (green circle). White filled circle indicates the sample mean and black filled circle indicates the mean predicted by the proposed method. Streamlines depict the flow field produced by the rotor control at the instant shown.}
    \label{fig:wall_stream}
\end{figure*}

This sort of flow structure seems to be typical of the optimal control solutions in this setting.  To verify this, we solve the control problem with the same parameters but with an initial density centered about $(-0.5,1)$.  That is, the initial density is $\rho(x(0)) = \N((-0.5,1),0.025I_2)$.  Fig. \ref{fig:cohsets_xm05y1} shows the $10^4$ data points colored according to the 3rd left singular vector of $A$. With both the initial distribution and the target in the left half of the domain, the second left singular vector simply divides the domain roughly into its left and right halves.  
However, for this case the third singular vector shows a partition which indicates a coherent structure that extends from the initial blob location at the initial time (see Fig. \ref{fig:cohsets_xm05y1} (a)) to the target at the final time (see Fig. \ref{fig:cohsets_xm05y1} (b)). 
Evaluating the objective in Eq. \ref{eq:mod_g} for this case, we have that $\bar{g}(f_X^2,f_Y^2) = 0.9978$ and $\bar{g}(f_X^3,f_Y^3) = 0.9928$ where $f_X^2,f_Y^2$ and $f_X^3, f_Y^3$ refer to the partition functions given by the second and third singular vectors, respectively. 

\vspace*{-0.15in}
\section{Transport by rotors near an infinite plane wall} \label{sec:wall}

For the case of a rotlet located at a point $\x_r$ above and infinite plane wall at $y=w$, the fluid flow must satisfy the additional boundary conditions of no slip and no penetration at the plane wall.  
The stream function associated with this flow is given by \cite{pozrikidis1992boundary,ranger1980eddies} 
\begin{equation}
\begin{split}
    \psi_w(\x) &= \gamma \Bigg(-\log|\x-\x_r| + \log|\x - \x_\im|  \\
    &\qquad \qquad - \frac{2 (y- w)( y-y_\im)}{r_\im^2} \Bigg)
\end{split}
\end{equation}
where $\x_\im = (x_\im,y_\im) = \left(x_r, 2w-y_r\right)$ is the location of an image singularity which has the effect of making the fluid velocity vanish at the plane wall.  Similarly, $r_\im = |\x - \x_\im|$.  Therefore, the flow produced for this case is $\mathbf{u}_w = (u_w,v_w)$ where 
\begin{subequations} 
\begin{align}
    u_w &= \gamma\Bigg(-\frac{y-y_r}{r^2} - \frac{y-y_\im}{r_\im^2} - \frac{2(y-w)}{r_\im^2} \\
    & \qquad \qquad + \frac{4(y-w)(y-y_\im)^2}{r_\im^4}\Bigg) \nonumber\\
    v_w &= \gamma\Bigg(\frac{x-x_r}{r^2} - \frac{x-x_r}{r_\im^2} \\
    & \qquad \qquad - \frac{4(y-w)(x-x_\im)(y-y_\im)}{r_\im^4} \Bigg) \nonumber 
\end{align}
\end{subequations}
and the velocity field for multiple rotlets above a plane wall can be found by summing the individual velocity fields as in Eq. \ref{eq:rotlet_vel}.

\begin{figure*}
    \centering
    \begin{minipage}{0.28\linewidth}
        \includegraphics[width = \linewidth]{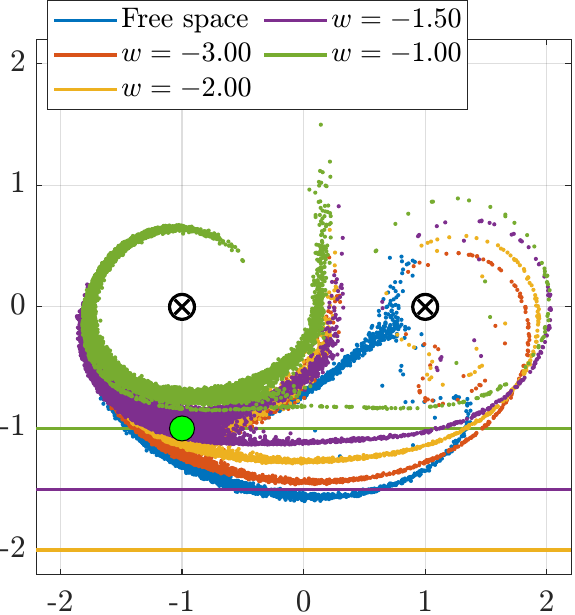}
        (a)
    \end{minipage}
    \begin{minipage}{0.35\linewidth}
        \vfill
        \includegraphics[width=\linewidth]{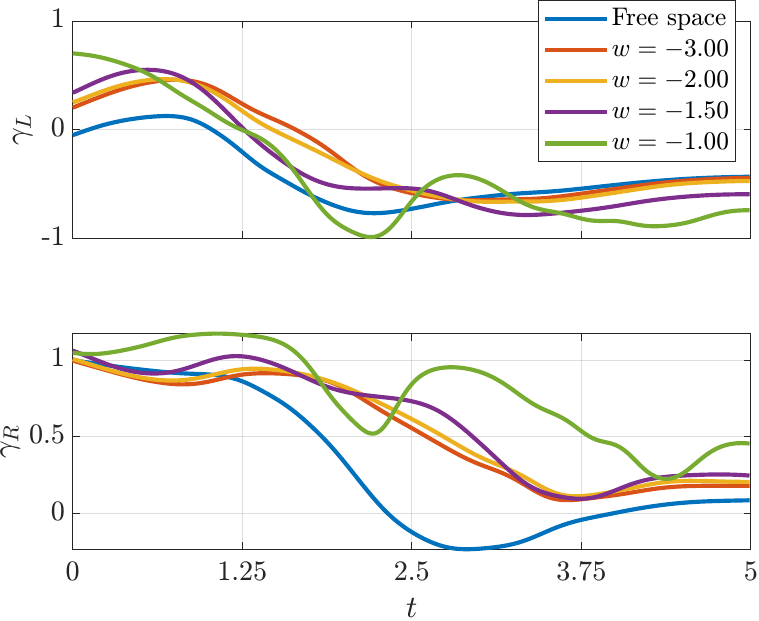}
        \vfill
        (b)
    \end{minipage}
    \begin{minipage}{0.33\linewidth}
        \vfill
        \includegraphics[width=\linewidth]{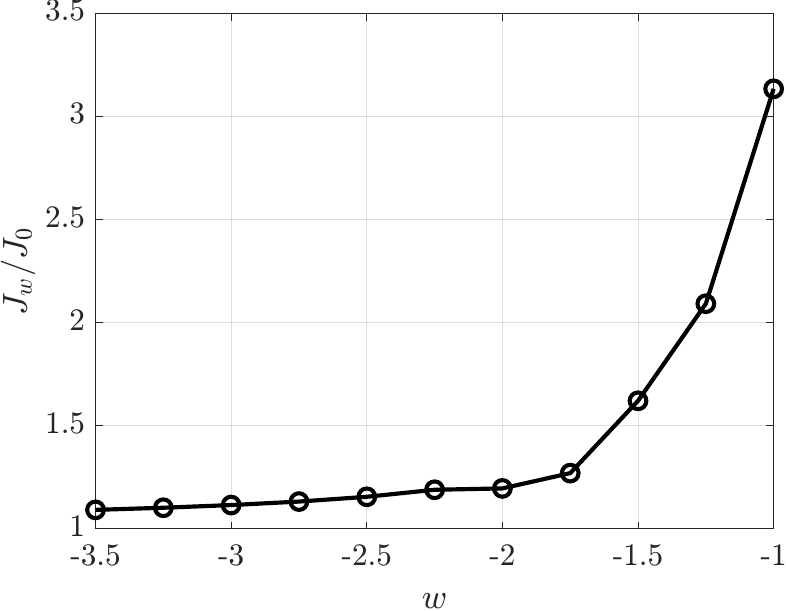}
        \vfill
        (c)
    \end{minipage}
    \caption{Comparison of boundary effects due to an infinite plane wall at various locations, $y=w$. (a) Overlaid final particle distributions at $t = 5$ for varying wall locations.  (b) Left and right rotor strengths as selected through the DDP algorithm for varying wall locations. (c) Total cost from the optimal control problem near a wall, $J_w$ normalized by the cost for the free space case, $J_0$. }
    \label{fig:wall_comp}
\end{figure*}

\begin{figure}[]
    \centering
    \includegraphics[width=\linewidth]{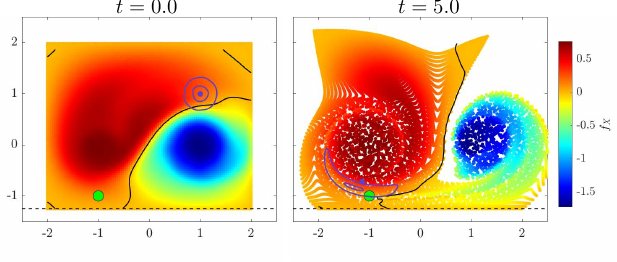}
    \vspace{-0.7cm}
    \caption{Finite-time coherent sets for the case shown in Fig. \ref{fig:wall_stream} with an infinite plane wall at $y=-1.25$.  Left: the initially uniform data set colored according to the partition function $f_X$, as approximated by the left singular vector of $A=\frac{1}{m}\Psi_X\Psi_Y^T$.  Right: The same points at the final time. The black line shows the $f_X=0$ level set. The purple marker and lines show the mean value and level sets of the density at values of the initial density at one and two standard deviations from the mean. }
    \label{fig:cohsets_wall}
    \vspace{-1em}
\end{figure}

With these governing equations for the flow produced by microrotors in the presence of a plane wall, we consider a similar transport problem to the one considered in Sec. \ref{sec:freespace} in order to examine the boundary effects of the plane wall on the transport problem. As in Sec. \ref{sec:freespace}, the same rotor positions of $(-1, 0)$ and $(1,0)$, initial density of $\rho(x(0)) = \N\big((1,1),0.025\,I_2\big)$, timespan of 5 units, target moments, and cost function are considered. The Liouville operators are computed using trajectory data from the same grid of initial conditions and basis functions positioned on the same grid as in Sec. \ref{sec:freespace}, but with any points in these grids lying outside of the fluid domain (below the plane wall) neglected. 

Fig. \ref{fig:wall_stream} shows the resulting flow field and its effect on the motion of the particle distribution for the case of a plane wall located at $w = -1.25$ from the control computed using the DDP algorithm.  From this figure, it is clear that the effect of the wall is to stretch the particle distribution along the wall due to the vanishing fluid velocity at the wall.  Due to this effect, the control tends to pull the distribution to the left in the early stages of the trajectory using a larger positive (counterclockwise) strength of the left rotor than in the free space case.  Related to this, in the middle stages of the trajectory, a larger positive strength of the right rotor is needed to supplement effects of the left rotor, as compared to the free space case.  These effects can also be clearly seen in Fig. \ref{fig:wall_comp} (b), which shows a time sequence of the rotor strengths for this problem for varying wall locations. Fig. \ref{fig:wall_comp} (a) shows an overlay of the final particle distribution at $t = 5$ for the same wall locations.  From this figure, it can be seen that effect of the wall is to elongate the distribution more for cases where the wall is closer to the target mean location.  Fig. \ref{fig:wall_comp} (c) shows a comparison of the optimal cost found from the DDP algorithm at varying wall locations, which demonstrates that the cost increases significantly as the wall nears the target mean position. This is due to both to the increased control effort (rotor strength) needed to steer the distribution as well as well as greater error in the moments due to the stretching effect of the wall.  Fig. \ref{fig:cohsets_wall} shows the coherent sets for the case shown in Fig. \ref{fig:wall_stream} at the initial and final times.  As in the free space case, the optimal control forms a coherent structure which passes near to the initial blob location at the initial time and extends toward the target at the final time.  

\begin{figure*}
    \includegraphics[width=\linewidth]{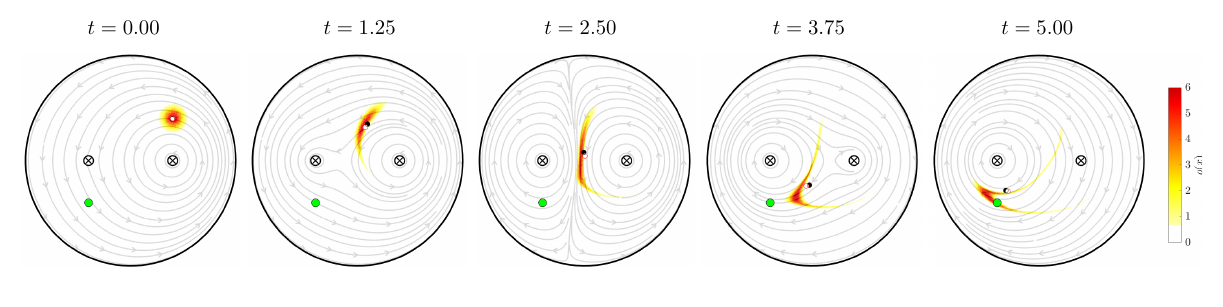}
    \vspace{-1cm}
    \caption{Controlled transport of a distribution of fluid particles within a circular boundary of radius $a = 2.5$ (black circle) by two micro-rotors fixed at $(-1,0)$ and $(1,0)$  from an initial density $\rho(x(0)) = \N((1,1),0.025I_2)$ to a target mean at $(-1,-1)$ (green circle). White filled circle indicates the sample mean and black filled circle indicates the mean predicted by the proposed method. Streamlines depict the flow field produced by the rotor control at the instant shown. }
    \label{fig:circle_stream}
\end{figure*}

\begin{figure*}
    \centering
    \begin{minipage}{0.33\linewidth}
        \includegraphics[width = \linewidth]{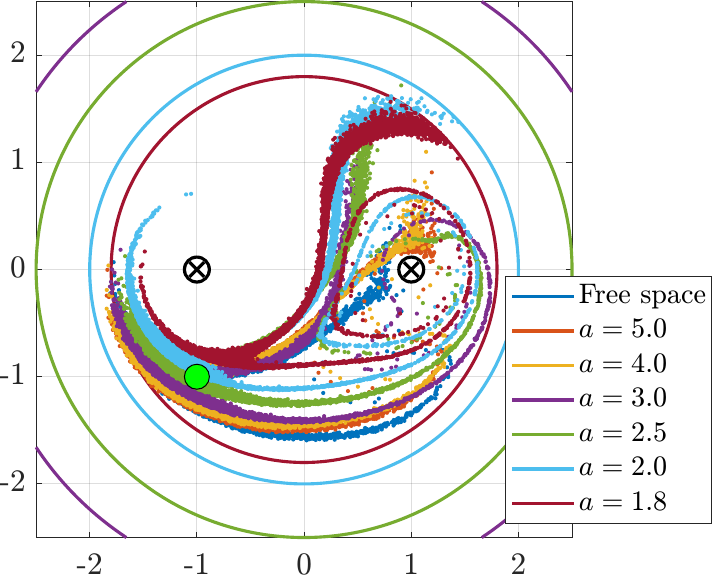}
        (a)
    \end{minipage}
    \begin{minipage}{0.35\linewidth}
        \vfill
        \includegraphics[width=\linewidth]{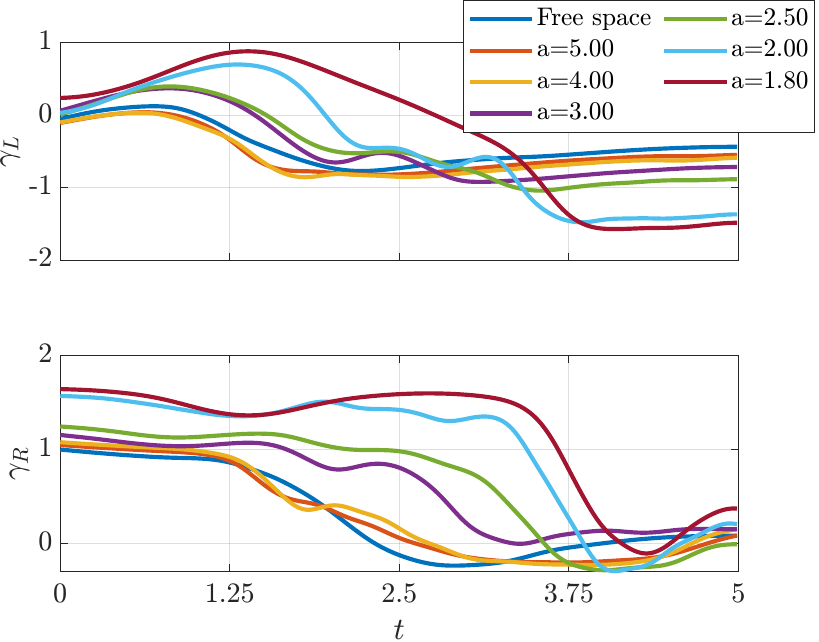}
        \vfill
        (b)
    \end{minipage}
    \begin{minipage}{0.28\linewidth}
        \vfill
        \includegraphics[width=\linewidth]{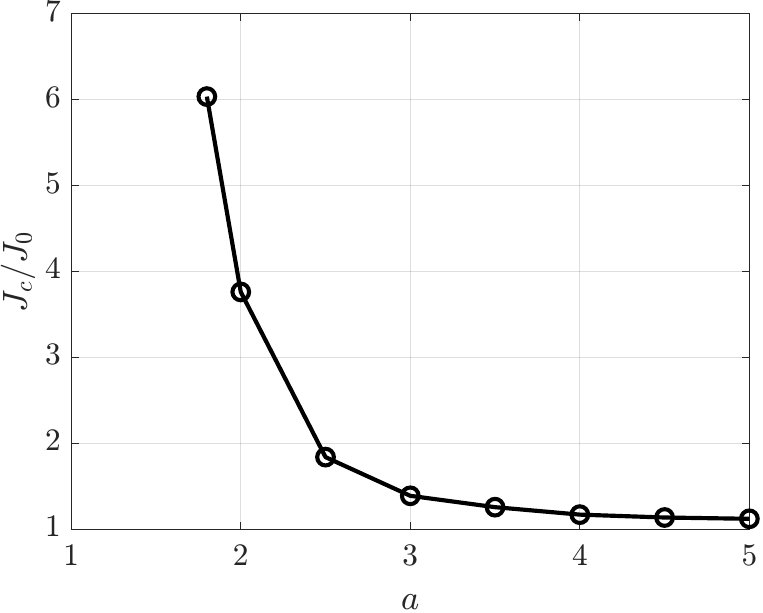}
        \vfill
        (c)
    \end{minipage}
    \caption{Comparison of boundary effects due to a circular boundary of varying radius, $a$. (a) Overlaid final particle distributions at $t = 5$ for varying boundary radius.  (b) Left and right rotor strengths as selected through the DDP algorithm for varying boundary radius. (c) Total cost from the optimal control problem in a circular boundary, $J_c$ normalized by the cost for the free space case, $J_0$. }
    \label{fig:circle_comp}
\end{figure*}

\section{Transport by rotors within a circular boundary}\label{sec:circle}
For the case of a rotlet positioned at a point $\x_r$ inside of a circular boundary of radius, $a$, centered about the origin, again the no-slip and no penetration boundary conditions must be satisfied by the flow at the boundary, and again, these can be satisfied by modifying the stream function to include image terms to cancel out the flow at the wall.  The stream function satisfying these conditions can be shown to be \cite{pozrikidis1992boundary,ranger1980eddies,arefmeleshko}
\begin{equation}
\begin{split}
    \psi_c = \gamma\Bigg(-\log|&\x-\x_r| + \log|\x-\x_\im| + \log \frac{R_r}{a} \\ 
    &- \frac{1}{2}\left(\frac{R^2-a^2}{x_\im^2}\right)\left(\frac{a^2}{R_r}-\frac{R^2}{a^2} \right) \Bigg)
\end{split}
\end{equation}
where $R = |\x|$ and $R_r = |\x_r|$ are the radial distances from the center of the circle to the evalutation point and to the rotlet, respectively, and $\x_\im = \frac{a^2}{R_r^2}\x_r$ is the location of the image system.  That is, the image is located outside of the circular boundary at a point along the line between the center of the circle and the rotlet at a radial distance of $a^2/R_r$ from the center of the circle.  
Then the flow field for this case is given by $\mathbf{u}_c = (u_c,v_c)$ where 
\begin{subequations}
\begin{align}
    u_c &= \gamma\Bigg(-\frac{y-y_r}{r^2} + \frac{y-y_\im}{r_\im^2} \\
   &\qquad \qquad + \left(\frac{a^2}{R_r^2} - 2\frac{R^2}{a^2}+1\right) \frac{y}{r_\im^2} \nonumber \\ 
   &\qquad \qquad - \frac{(y-y_\im)(R^2-a^2)\left(\frac{a^2}{R_r}-\frac{R^2}{a^2}\right)}{r_\im^4} \Bigg) \nonumber \\
   v_c &= \gamma\Bigg(\frac{x-x_r}{r^2} - \frac{x-x_\im}{r_\im^2} \\
   &\qquad \qquad - \left(\frac{a^2}{R_r^2} - 2\frac{R^2}{a^2}+1\right) \frac{x}{r_\im^2} \nonumber\\ 
   & \qquad \qquad+ \frac{(x-x_\im)(R^2-a^2)\left(\frac{a^2}{R_r}-\frac{R^2}{a^2}\right)}{r_\im^4}\Bigg) .\nonumber
\end{align}
\end{subequations}

With these governing equations for the flow produced by microrotors within a circular boundary, we consider the same transport problem considered in previous cases in order to examine the boundary effects of the circular boundary on the transport problem. As before, the rotor positions of $(-1, 0)$ and $(1,0)$, initial density of $\rho(x(0)) = \N\big((1,1),0.025\,I_2\big)$, timespan of 5 units, the same target moments and cost function are considered. The Liouville operators are computed using trajectory data from the same grid of initial conditions and basis functions positioned on the same grid as in Sec. \ref{sec:freespace}, but with any points in these grids lying outside of the fluid domain (beyond the circular boundary) neglected.

\begin{figure}
    \centering
    \includegraphics[width=\linewidth]{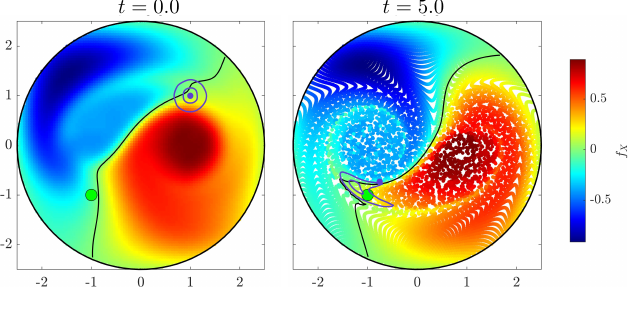}
    \caption{Finite-time coherent sets for the case shown in Fig. \ref{fig:circle_stream} with a circular boundary of radius $a=2.5$.  Left: the initially uniform data set colored according to the partition function $f_X$, as approximated by the left singular vector of $A=\frac{1}{m}\Psi_X\Psi_Y^T$.  Right: The same points at the final time. The black line shows the $f_X=0$ level set. The purple marker and lines show the mean value and level sets of the density at values of the initial density at one and two standard deviations from the mean.}
    \label{fig:cohsets_circle}
\end{figure}

Fig. \ref{fig:circle_stream} shows the resulting flow field from the control and its effect on the motion of the particle distribution for the case of the two rotors within a circular boundary of radius $a = 2.5$.  
Similarly to the case next to a plane wall, the reduced fluid velocity near the circular boundary leads to a stretching effect on the distribution, especially when a significant part of the particle distribution lies in regions near to the boundary.
Since this is encountered at the initial condition, significantly more particles remain in the upper, trailing `tail' of the distribution due to the drag effects of the boundary in the upper right quadrant. 
This effect becomes more apparent for smaller boundary radius. 
Due to this effect, more control effort must be exerted by the rotors in the early stages of the trajectory to overcome this drag.  
A secondary effect of this is that the leading tail of the distribution, which consists of particles closer to the interior of the circle and further from the boundary, tends to stretch more, leading it to wrap around the rightmost rotor in the later stages of the trajectory in a way that was not seen in the previous cases. 
These qualitative differences are highlighted in Fig. \ref{fig:circle_comp} (a), which shows an overlay of the final particle distribution at $t = 5$ for the same boundary radius.
Fig. \ref{fig:circle_comp} (b) shows a time sequence of the rotor strengths for this problem for varying wall locations.
Fig. \ref{fig:cohsets_circle} shows the coherent sets for the case shown in Fig. \ref{fig:circle_stream} at the initial and final times.  As in the previous cases, the optimal control produces a flow field a coherent structure which passes near to the initial blob location at the initial time and extends toward the target at the final time.

\section{Transport of two densities} \label{sec:two_blobs}

We now return to the case of two micro-rotors in free space to consider the problem of manipulating two distinct distributions of fluid particles to a common target mean and second moment.  This requires a reformulation of the optimal control problem as posed in Eq. \ref{eq:optimization}. In that formulation, the state of the control problem was taken to be the vector of projection coefficients $\hat{\rho}$.  Here we consider an augmented state containing the projection coefficients of the two density functions. Denoting these two density functions as $\rho^A$ and $\rho^B$, and their corresponding projection coefficients by $\rh^A$ and $\rh^B$, the augmented state for this case is $[(\rh^A)^T, (\rh^B)^T]^T$. Similarly, we consider an output vector which concatenates the first and second moments for the two density functions $y = [(m^A)^T , (m^B)^T]^T$, where $m^A$ and $m^B$ are vectors containing the moments of the densities $\rho^A$ and $\rho^B$ respectively, as in Eq. \ref{eq:momentvec}.  The same Liouville operators are used to propagate each of these densities forward in time.  From this point, an appropriate cost function can be specified and the rotor control can be optimized using the DDP scheme as before.  

With this formulation, we consider the problem of manipulating two densities using two rotors fixed at the same locations as before, $(-1,0)$ and $(1,0)$.  We take the initial density for one of the distributions to be the same as the previous examples, $\rho^A(x(0)) =  \N\big((1,1),0.025\,I_2\big)$, and consider a second distribution starting from an initial density of $\rho^B(x(0)) = \N\big((b,1),0.025\,I_2\big)$, where $b$ is a parameter to be varied.  This formulation allows us to examine the ability to steer two distributions starting from varying initial distances apart.  
We consider the problem of choosing the rotor strengths to steer both of these distributions to a final distribution with a mean of $m_1^A(5) = m_1^B(5) = (-1,-1)$, while minimizing the variances.
For this problem, the cost function is taken to be of the same form as Eq. \ref{eq:cost_both} with the weights chosen to be $S = 0.1I_{10}$, $R = I_2$, and $S_H = 500I_{10}$.  That is, the terminal cost is chosen to be half that of the previous cases since it is being applied to the error in the moments of two density functions and summed.   

\begin{figure}[]
    \centering
    \includegraphics[width = \linewidth]{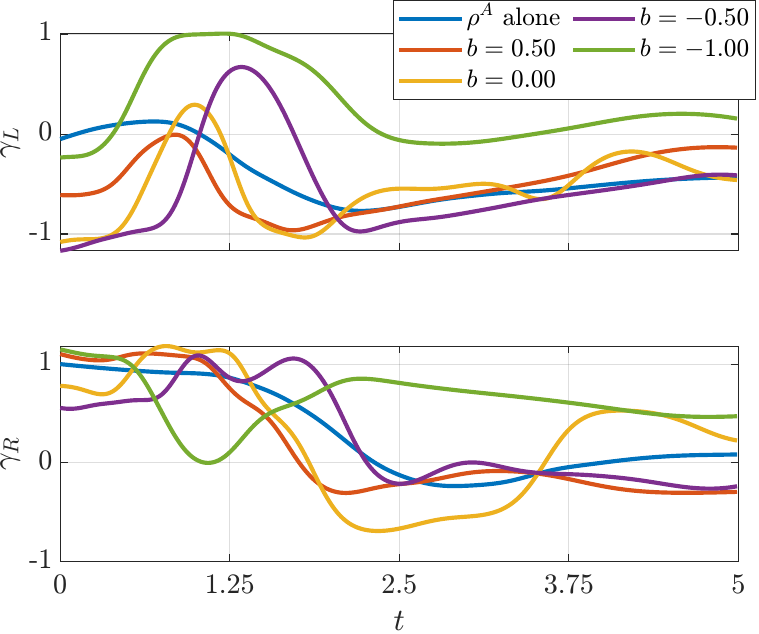}
    \caption{Rotor strengths for controlled transport for varying initial distributions, $\rho^B(x(0)) = \N((b,1),0.025I_2)$ with $\rho^A(x(0)) = \N((1,1),0.025I_2)$, corresponding to the sequences shown in Fig. \ref{fig:mix_stream}. }
    \label{fig:mix_control}
\end{figure}

\begin{figure*}
\includegraphics[width=\linewidth]{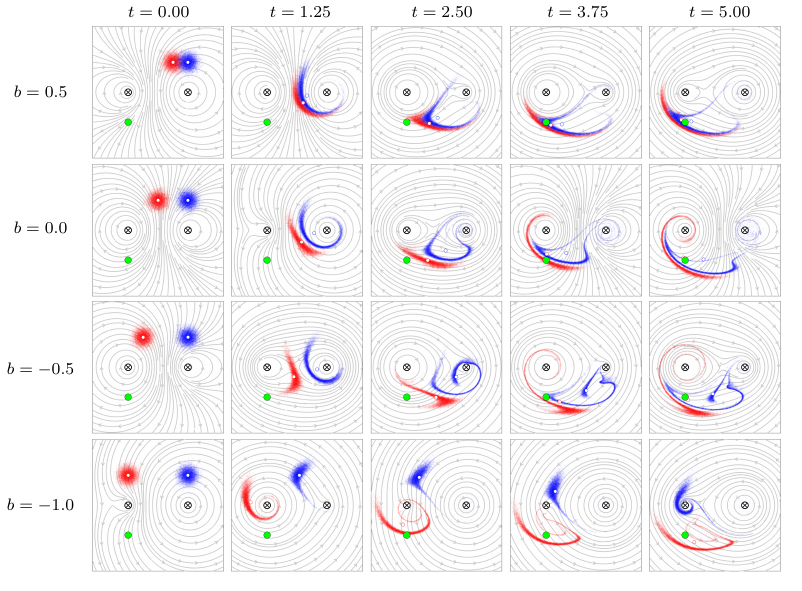}
    \caption{Controlled transport of two distributions of fluid particles by two micro-rotors fixed at $(-1,0)$ and $(1,0)$  from initial densities $\rho^A(x(0)) = \N((1,1),0.025I_2)$ and $\rho^B(x(0)) = \N((b,1),0.025I_2)$ to a target mean at $(-1,-1)$ (green circle).  Each row shows the results for a different value of $b$: first row: $b = 0.5$, second row: $b = 0$, third row: $b = -0.5$, fourth row: $b = -1$. White filled circles indicate the sample means of the two distributions. Streamlines depict the flow field produced by the rotor control at the instant shown. }
    \label{fig:mix_stream}
\end{figure*}

Fig. \ref{fig:mix_stream} shows snapshots from the evolution of the particle distributions for the flow induced by the rotors controlled using the strengths determined from the DDP algorithm for four different initial distributions $\rho^B$ with the initial $x$-coordinate of the mean being $b = 0.5$, $b = 0.0$, $b=-0.5$, and $b=-1.0$, respectively, on the rows.   It can be seen that the flow produced in the case where $b = 0.5$ is qualitatively similar to the case of controlling the density $\rho^A$ alone, as was considered in Sec. \ref{sec:freespace}.  For the next two cases of $b = 0$ and $b = -0.5$, we see that as the initial distribution of $\rho^B$ starts farther from the initial distribution of $\rho^A$, a higher negative spin is applied by the left rotor in the early stages of the trajectory, producing a flow that is more symmetric as the blobs are pulled toward the middle, but with a similar flow near the end of the trajectory as the distributions near the target.  In the last case shown, where $b = -1.0$, it appears that a transition has occurred and a qualitatively different optimal trajectory is found in which the leftmost distribution $\rho^B$ is stirred counterclockwise around the left rotor rather than through the region between the rotors.  This is done by a positive torque applied from the left rotor, which also results in the rightmost blob $\rho^A$ being pulled to a position above the left rotor.  As a result of this, at the end of the sequence, the rightmost rotor generates a counterclockwise flow which pushes the two distributions down toward the target.  This is in contrast to the other cases, where the right rotor generates a clockwise flow near the end in order to steer the particles from right to left toward the target. These effects can also be seen by examining the rotor strengths directly, as shown in Fig. \ref{fig:mix_control}.

\begin{figure}
    \centering
    \includegraphics[width = \linewidth]{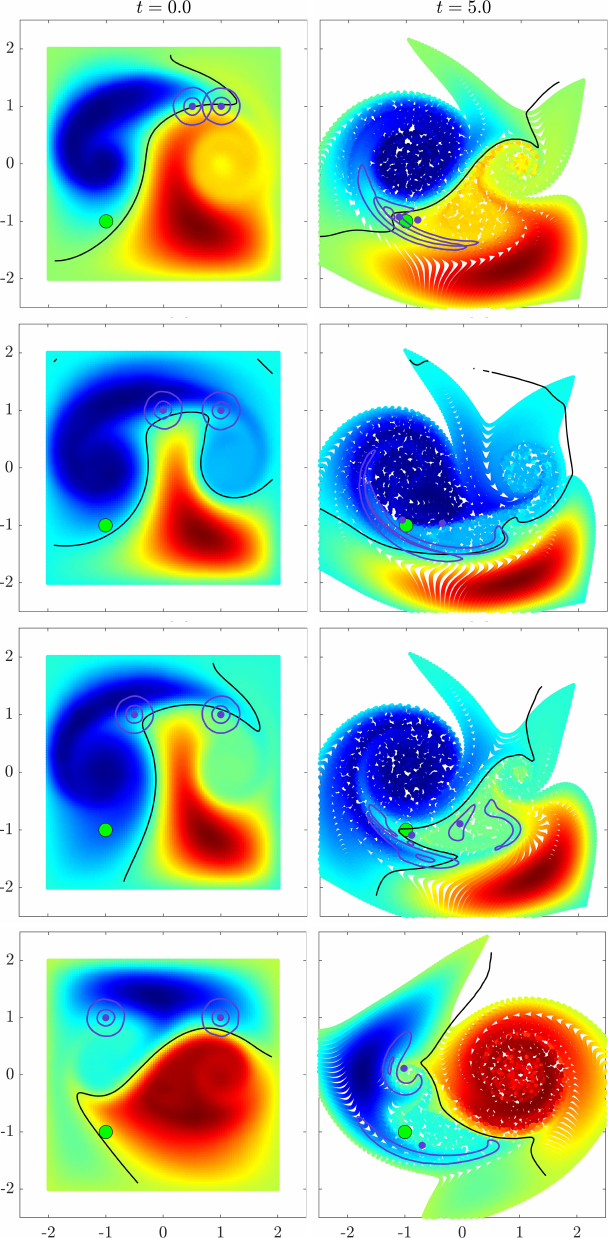}
    \caption{Finite-time coherent sets for the the transport of two distributions, as shown in Fig. \ref{fig:mix_stream}.  Left column: the initially uniform data set colored according to the partition function $f_X$, as approximated by the left singular vector of $A=\frac{1}{m}\Psi_X\Psi_Y^T$.  Right: The same points at the final time. The black line shows the $f_X=0$ level set. The purple marker and lines show the mean value and level sets of the density at values of the initial density at one and two standard deviations from the mean.  Each row shows the results for a different value of $b$: first row: $b = 0.5$, second row: $b = 0$, third row: $b = -0.5$, fourth row: $b = -1$.}
    \label{fig:cohsets_mix}
\end{figure}

Fig. \ref{fig:cohsets_mix} shows the coherent sets at the initial and final time for the cases shown in Fig. \ref{fig:mix_stream} where two distributions are to be steered to the common target.  In the first three cases considered, the coherent structure which divides the coherent sets at the initial time passes through the regions of high concentration of both initial distributions.  At the final time, this structure moves toward the target, effectively pulling both distributions toward the goal. 

\section{Conclusion}
A promising new approach has been developed and demonstrated for computing the optimal control to transport a distribution of states whose dynamics are governed by a control affine system to a desired final state distribution in a fixed, finite time.  We demonstrate the usefulness of this method by highlighting a fluid mechanical application, in which the relevant state is the position of a fluid particle,  the distribution describes a blob of fluid particles, and the controls are the torques applied by a pair of fixed rotors, which stir the flow in circular patterns. 
In this setting, we used the proposed approach to analyze the effects of fixed boundaries on the transport problem.
We believe that such control strategies will be very useful in applications, such as targeted drug delivery, particle manipulation, and cell sorting in which the relevant transport problem is not to mix the fluid, but to transport a concentrated distribution of particles in a controlled way to a desired location. In future works, we plan to study similar transport problems in which the flow is generated by non-stationary stirrers, such as a moving rotors or microswimming robots\cite{buzhardt2019dynamics}, or by boundary controls. 
Other interesting use case of the work presented here could be using this algorithm to optimize rotor placement for a given task.  This application could be especially relevant for the design of microfluidic devices where fluid transport is critical.  

While it was demonstrated on and motivated by problems in the fluids setting, we believe that the proposed approach can have much broader application in control systems, where the density of states can be taken to represent an uncertainty distribution \cite{chen2021controlling}.  
Other exciting extensions of this work could include understanding the relationship between this method and the formation, motion, and manipulation of transport barriers in a flow field.

\bibliography{rotlets}

\end{document}